# Numerical Analysis of High-index Nano-composite Encapsulant for Light-emitting Diodes

**Young-Gu Ju**

**KyungPook National University, Sankyuk, Daegu, 702-701 Korea**

**Guilhem Almuneau**

**LAAS-CNRS, 7, Avenue du Colonel Roche F-31077 Toulouse Cedex 4 France**

**Tae-Hoon Kim**

**Korea Photonics Technology Institute, Wolchul dong 971-35, Gwangju, 500-460 Korea**

**Baek-Woon Lee**

**LCD Business Unit, Samsung Electronics Corporation, San 24, Nongseo-ri, Giheung-eup, Yongin-si, Gyeonggi-do, 449-901 Korea**

We used two-dimensional Finte-Difference-Time-Domain (FDTD) software to study the transition behavior of nano-particles from scatterers to an optically uniform medium. We measured the transmission efficiency of the dipole source, which is located in the high refractive index medium(index=2.00) and encapsulated by low index resin(index=1.41). In an effort to compose index-matched resin and to reduce internal reflection, high-index nano-particles are added to low-index resin in simulations of various sizes and densities. As the size of the nano-particles and the average spacing between particles are reduced to 0.02 λ and 0.07 λ respectively, the transmission efficiency improves two-fold compared to that without nano-particles. The numerical results can be used to understand the optical behavior of nano-particles and to improve the extraction efficiency of high brightness light-emitting-diodes(LEDs), through the use of nano-composite encapsulant.

Email: ygju@knu.ac.kr

Fax: +82-53-950-6893

# I. INTRODUCTION

GaN-based light-emitting diodes(LEDs) have attracted considerable attention in relation to lighting applications. As brightness has increased, applications such as displays, traffic signals, and backlights for cellular phone have become possible. With these trends, applications may extend into backlights for liquid crystal display televisions and even the general illumination industry in the near future. In order to enter the general lighting market, LEDs must improve the light output and luminous efficacy significantly. The lighting industry demands LEDs that can attain a luminous efficacy in excess of 150 lm/W, at a cost of less than 5 $/klm to replace fluorescent lamps [1]. Improvements are being tested at several levels: Internal quantum efficiency [2-3], extraction efficiency [4-9], and phosphor-conversion efficiency [10]. Among the various research, herein, we focus on improving extraction efficiency.

The external quantum efficiency of the GaN-based LEDs is low because the refractive index(RI) of the nitride epitaxial layer differs greatly from that of air. The critical angle at which light can escape is only about 23 degrees. Some [4-9] have taken on the challenge of exploring new shaped chips, photonic crystals and roughened surfaces. Although all of these methods have improved efficiency, they require some extra processing steps in the device level. Especially, if the processing requires high-resolution lithography, as in the case of photonic crystals [4], it may raise production costs to an unacceptable level. As an alternative to the modification of the device surface, an attempt to find index-matched resin in the packaging can be considered. In general, resin that is used as a lens in LED packaging has an index of 1.5, whereas the index of the GaN epitaxial layer is 2.0. The mismatch between the GaN epitaxial layer and resin still induces a considerable internal reflection and allows only a small escape angle. If resin with an index of 2.0 can be composed, the extraction efficiency will greatly improve. In fact, this approach has already been tried to prevent the scattering of light at the interface between resin and phosphor particles [10]. They obtained high-index resin by adding nano-particles to the host matrix. The achieved index is 1.8, which is the same as the index of the bulk phosphor. From the high-index approach, scattering could be reduced significantly, and an enhancement of 10-20 % in the lumen output could be observed, as compared to the case of conventional encapsulant(RI-1.5). In spite of the great success in the laboratory, it is relatively difficult to find the theoretical background for the optical effect of nano-composite in literature. Without a theoretical background, the optimization process for the search of high-index resins can be a series of time-consuming experiments.

As matter of fact, the basic optical theory behind nano-composites can be found in most textbooks dealing with the fundamentals of optics [11]. The origin of index engineering can be traced to dipole oscillation or the scattering phenomenon. Textbooks, however, generally mention the scattering of single particle with wavelength sizes or less. If the size shrinks to atomic size, the phenomenon can be approximated by dipole oscillation or by the theory of a uniform medium. The physics regarding the transition from the scattering of large particles to the transparent medium has not been explained well, even though this transition region is the main playground where nano-particles function. In this paper, we tackle this problem using the Finite-Difference-Time-Domain(FDTD) and investigate the optical behavior of nano-particles near the transition region, in order to provide some meaningful numbers for the engineering of high-index nano-composite encapsulant for LEDs.

## II. NUMERICAL MODELING

As mentioned before, we employed the FDTD to study the optical transition behavior of nano-particles as sizes decrease. FDTD is basically a method that solves the partial differential equation by meshing space into small grids and applying the local interaction law, specified by the Maxwell equation, to those nodal points. In a simulation, we used two-dimensional(2D) FDTD in order to save computation time and scan a wide parameter range. There is no doubt that a three-dimensional(3D) program provides more exact numerical data, but it also requires more calculation time, as compared to the 2D program. The purpose of this paper is to observe the transition behavior of nano-particles for qualitative descriptions and to provide some meaningful numbers for the engineering of a nano-composite. More accurate numbers using 3D FDTD will be the target for the future research.

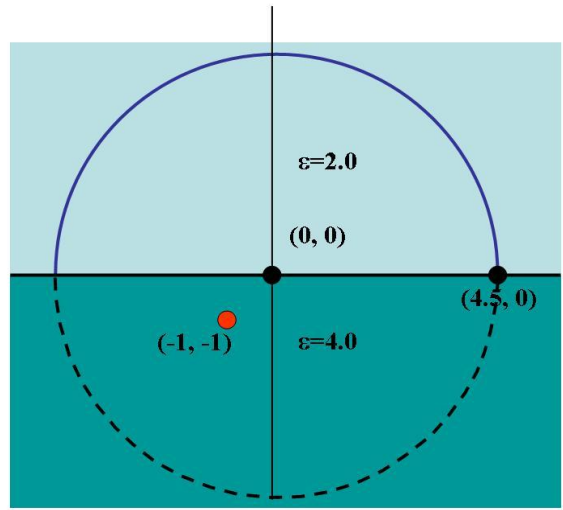

**Fig. 1 Schematic diagram of a simulation setup. The permittivities of the upper and lower half regions are 2.0 and 4.0, respectively. The dipole source is located at (-1, -1) while the integration of the energy flux for the upper region is carried out along the half circle(blue solid line) with a radius of 4.5.**

The situation used for a simulation is illustrated in Fig. 1. A space is divided into two sections by a horizontal boundary. The upper half has a lower index than the lower part. The dipole source is placed at (-1, -1), which leads to internal reflection at a critical angle of approximately 45 degrees. The dipole source is TE(Transverse Electric)-polarized unless otherwise mentioned. In this paper, TE polarization refers to the case in which the magnetic field is perpendicular to the plane of incidence. The index of the lower part is close to that of the GaN epitaxial layer and the index of the upper part is also similar to that of the epoxy lens. Measurements are taken by integrating the Poynting vector along the half circle(solid line) in the upper medium. Only the normal component to the half circle is considered. The total amount of energy flux is checked by integrating the Poynting vector along the full circle(the sum of the solid and dotted lines). The total energy flux is independent of the radius of the circle and the position of dipole source. Actually, the total amount of energy shows some fluctuation when the dipole is near the horizontal boundary. The coordinates (-1, -1) are far from the horizontal interface so as to satisfy energy conservation laws. Considering that all lights don't come from the exact center of the chip and epoxy lens, the positioning of the dipole source off the center can be justified. If the surface of the lens follow the line of the detection cup(the upper half circle) with an anti-reflection coating, the emission efficiency will be the ratio of the output power measured at the detection cup, to the total output power of the dipole.

Nano-particles are implemented by inserting circular objects with a permittivity of 9.0 into the upper half region. The distribution of nano-particles in the host matrix is based upon a triangular lattice that prepares for a high packing ratio. Since nano-composites, in a real situation, are more similar to an amorphous structure than a crystalline structure, distribution needs some random factors in terms of the size and positioning of the nano-particles. Therefore, the randomness is realized in the modeling in order to avoid side effects or peculiarities that stem from periodicity. The density of the nano-particles is controlled by varying the triangular lattice constant. One example of a permittivity profile is shown in Fig. 2. In the configuration and the following calculation, a wavelength of $\lambda$ is used as a basic unit of length.

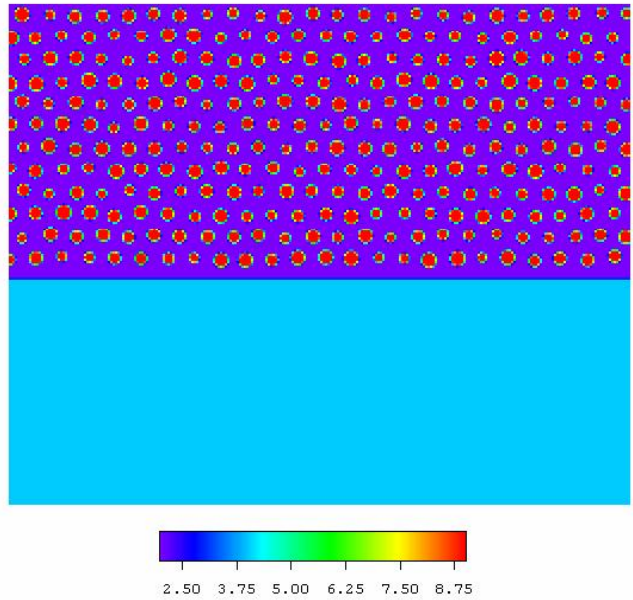

**Fig. 2 A permittivity profile used in the simulation. The average radius of the nano-particles ($r_{av}$) equals 0.02 $\lambda$. The average distance between particles($a_{av}$) equals 0.11 $\lambda$.**

## III. RESULTS

Unlike the conventional FDTDs[12, 13], we used nearly monochromatic light source by extending the decay time of the dipole source. The dipole source of FDTD usually generates a short pulse and the probe measures the variations of an electro-magnetic field. The nano-composite, however, should not be sensitive to small variation of frequency, since the mixture is expected to behave like an amorphous uniform material without the structural influence. As a result of the calculations, we can obtain field profiles as seen in Fig. 3-(a), (b) and (c). The first picture represents the field intensity observed in the reference setup, where no nano-particles are added to the upper region. The enhanced intensity of the field in the lower medium(RI=4.0), rather than in the upper medium(RI=2.0) is found along the boundary of the two media. The higher intensity of the field corresponds to the internal reflection of the dipole source at the interface, as anticipated. The second figure(Fig. 3-(b)) presents results when 0.05 $\lambda$ particles are added to the host matrix with the average lattice constant $a_{av}$ of 0.28 $\lambda$. At this composition, the differences from Fig. 3-(a) are hardly seen. The field profile still exhibits the strong internal reflection pattern along the interface. The situation, however, changes greatly as the size of the particles decrease and the density of the particles increase in the upper medium. Fig. 3-(c) shows a case in which $r_{av}$ is reduced to 0.02 $\lambda$ and the $a_{av}$ is 0.07 $\lambda$. The reduction in the internal reflection is clearly seen, as compared with the former two pictures.

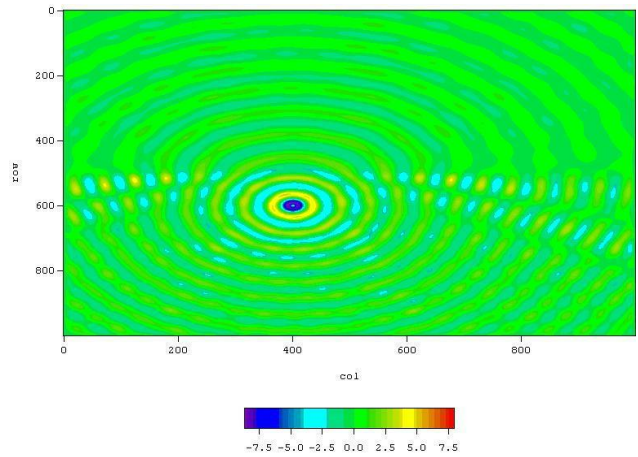

(a)

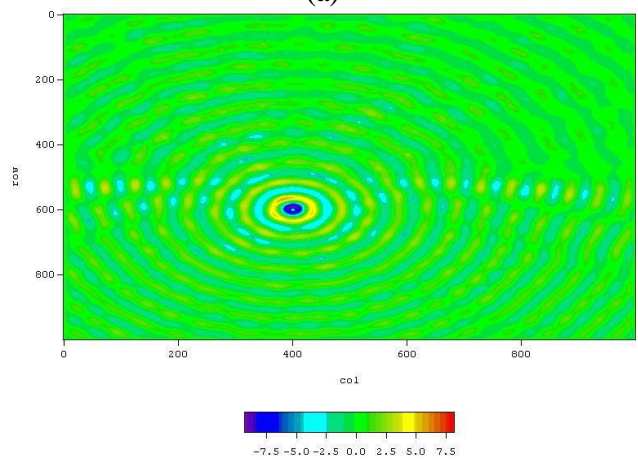

(b)

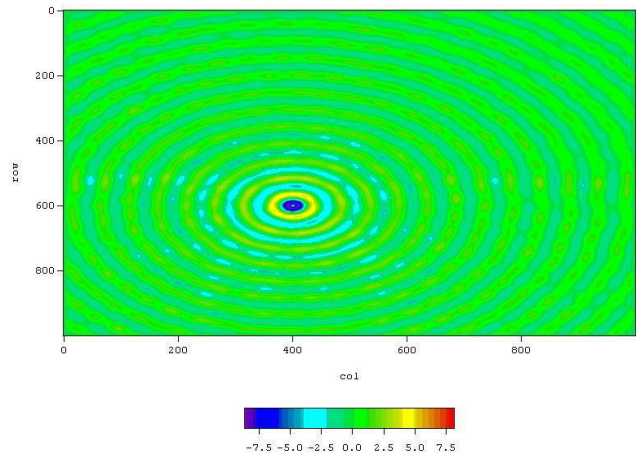

(c)

**Fig 3. Field intensity propagating from the dipole source through the medium. (a) No nano-particles, simple boundary consisting of RI=4.0/2.0. (b) $r_{av}$ =0.05 $\lambda$, $a_{av}$ =0.28 $\lambda$  (c) $r_{av}$ =0.02 $\lambda$, $a_{av}$ =0.07 $\lambda$**

Although the observation of the field intensity profile provides quite useful information in understanding the behavior of nano-composite qualitatively, simple observation is not sufficient for confirming the clearance of a scattering effect or the study of transition behavior. For a quantitative analysis, we need to define the transmission efficiency as the ratio of the output power that escapes the upper detection circle to the total output power, as mentioned in modeling and in Fig. 2. As the size and density of the nano-particles vary, the transmission efficiency is measured. In this way, we can compare the various structures and study the transition behavior numerically.

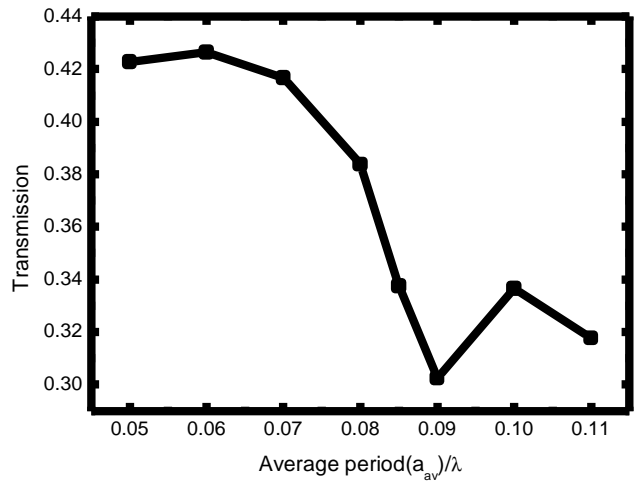

**Fig. 4. Transmission efficiency as a function of the average period($a_{av}$). $r_{av}$ =0.02 $\lambda$**

Fig. 4 shows how the transmission efficiency changes depending upon the average spacing between nano-particles. As expected, a lower efficiency is obtained at large distances, where the density is lower. At a low particle density, the effective index gets close to that of the host matrix(RI=1.41). The fluctuation at a large period may be attributed to the scattering and non-uniformity of the medium. As the average spacing decreases, the transmission approaches the ideal value. The ideal transmission efficiency of 0.44 is obtained when the uniform medium, with the same index, is assumed for both the upper and lower-half regions. On the contrary, the lowest transmission is 0.26, in the case where the medium is divided into two uniform regions, the indices are 2.00 and 1.41, respectively. With a period of 0.07 $\lambda$, the transmission efficiency is already close to peak value. In fact, the 0.07 $\lambda$ is the value at which the average index becomes 2.0 from the simple calculation with the weighting factor of area ratio. Therefore, if the period is decreased to less than 0.07 $\lambda$, the effective index can be higher than 2.0, thus, leading to a decrease in the transmission. Fig. 4, however, doesn't show such a roll-over, but only saturation near the maximum. There could be a discrepancy between a simple average index and the simulation based on Maxwell equation.

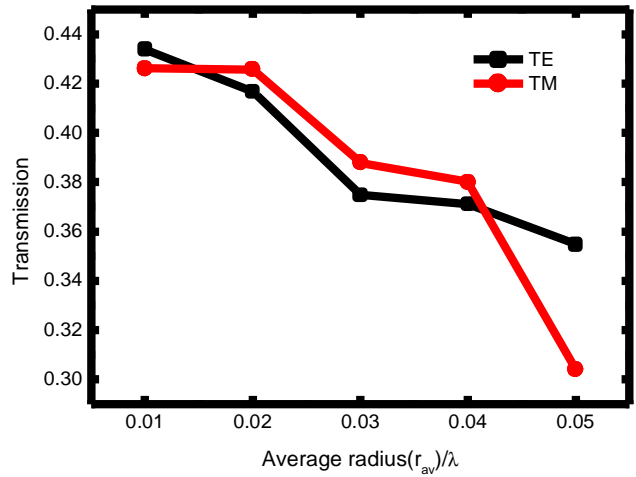

**Fig. 5. Transmission efficiency as a function of the average radius of nano-particles($r_{av}$). $a_{av}$ is adjusted at every point so that the simple average index is 2.0.**

The transition of a nano-particle from a scatterer to an atomic-like dipole in a uniform medium can be dramatically observed irrespective of polarization, as the radius of a nano-particle decreases. Fig. 5 shows how the transmission efficiency can be improved with the reduction in particle size. When the radius reaches 0.01 λ, it is already close to the ideal value. At this size, the nano-particles don't work as scatterers, which means the medium become transparent optically. Assuming that the wavelength is 400 nm, 0.01 λ is about 4 nm. In reference [10], the used nano-particles are in the size range of 7-12 nm. Therefore, a diameter of 8 nm in the simulation agrees with the experimental results. Polarization dependency can be partially ascribed to Fresnel coefficient of reflection, as normally observed at the interface between two uniform materials. In a uniform medium, TE polarization corresponding to p-polarization has a higher transmission at a large incident angle than TM(Transverse Magnetic) polarization, which corresponds to s-polarization.

## IV. CONCLUSION

In summary, the transition of nano-particles from scatterers to an optically smooth medium can be analysed using the 2D FDTD. The transition occurs when the size of a nano-particle is approximately 0.02 λ in diameter. The transmission efficiency of a nano-composite also depends on the density of nano-particles in the resin. It, however, may not follow the trends at a high density as expected from a simple index average calculation. Theoretically, we can improve the extraction efficiency of high brightness LEDs by controlling nano-particle size and density.

## ACKNOWLEDGEMENT

This work was supported by Regional Research Center for Photonic Materials and Devices at Chonnam National University under Grant R12-200-054 of Korea Institute of Industrial Technology Evolution and Planning. This work was also partially supported by funding from CNRS and KOSEF. We appreciate the help of Se-Heon Kim at KAIST who provided valuable advice, as well as the FDTD codes used in the simulation.

## REREFENCES

[1] J. Y. Tsao, Laser Focus World, S11-S14 (May 2003)

[2] M. R. Krames, J. Bhat, D. Collins, N. F. Gardner, W. Goetz, C. H. Lowery, M. Ludowise, P. S. Martin, G. Mueller, R. Mueller-Mach, S. Rudaz, D. A. Steigerwald, S. A. Stockman, and J. J. Wierer, Phys. Stat. Sol. (a) **192**, 237-245 (2002).

[3] T. Onuma, S. F. Chichibu, A. Uedono, T. Sota, P. Cantu, T. M. Katona, J. F. Keading, S. Keller, U. K. Mishra, S. Nakamura, and S. P. DenBaars, J. Appl. Phys. **95**, 2495-2504 (2004).

[4] T. N. Oder, K. H. Kim, J. Y. Lin, and H. X. Jiang, Appl. Phys. Lett. **84**, 466-468 (2004).

[5] M. R. Krames, M. Ochiai-Holcomb, G. E. Hofler, C. Carter-Coman, E. I. Chen, I. H. Tan, P. Grillot, N. F. Gardner, H. C. Chui, J. W. Huang, S. A. Stockman, F. A. Kish, M. G. Craford, T. S. Tan, C. P. Kocot, M. Hueschen, J. Posselt, B. Loh, G. Sasser, D. Collins, Appl. Phys. Lett. **75**, 2365-2367 (1999).

[6] T. Fujii, Y. Gao, R. Sharma, E. L. Hu, S. P. DenBaars, and S. Nakamura, Appl. Phys. Lett. **84**, 855-857 (2004).

[7] T. Gessmann, E. F. Schubert, J. W. Graff, K. Streubel, and C. Karnutsch, IEEE Electron. Device Lett. **24**, 683-685 (2003).

[8] Hung-Wen Huang, C. C. Kao, J. T. Chu, H. C. Kuo, S. C. Wang, and C. C. Yu, IEEE Photonics Technology Letters, **17**, 983-985 (2005)

[9] N. Narendran, Y. Gu, J. P. Freyssinier-Nova, and Y. Zhu, Phys. Stat. Sol. (a) **202**, R60-R62, (2005)

[10] N. Taskar, R. Bhargava, J. Barone, V. Chhabra, V. Chabra, D. Dorman, A. Ekimov, S. Herko, and B. Kulkarni, Proc. SPIE **5187**, 133-141 (2004).

[11] E. Hecht, *Optics*, 4$^{th}$ Ed. , (Addison Wesley, 2001), Chap. 4

[12] Se-Heon Kim, Han-Youl Ryu, Hong-Gyu Park, Guk-Hyun Kim, Yong-Seok Choi, Yong-Hee Lee, and Jeong-Soo Kim, Appl. Phys. Lett. , **81**, 2499-2501 (2002)

[13] Yong-Jae Lee, Se-Heon Kim, Guk-Hyun Kim, Yong-Hee Lee, Sang-Hwan Cho, Young-Woo Song, Yoon-Chang Kim, and Uoung Rae Do, "Far-field radiation of photonic crystal organic light-emitting diode" Optics Express, **13**, 5864-5870 (2005).